\documentclass{osa-article}

\usepackage{soul}
\usepackage{xcolor}
\usepackage{graphics}
\journal{oe}
\renewcommand{\floatpagefraction}{.8}

\begin{document}

\title{A self referencing attosecond interferometer with zeptosecond precision}

\author{Jan Tro\ss,\authormark{1}, Georgios Kolliopoulos\authormark{1,2}, and Carlos A. Trallero-Herrero\authormark{1,3,*}}

\address{\authormark{1}J.R. Macdonald Laboratory, Department of Physics, Kansas State University, Manhattan, Kansas 66506, USA\\
\authormark{2}Extreme Light Infrastructure - Nuclear Physics, Romania\\
\authormark{3}Department of Physics, University of Connecticut, Storrs, CT 06268, USA}
\email{\authormark{*}carlos.trallero@uconn.edu}



\begin{abstract}
In this work we demonstrate the generation of two intense, ultrafast laser pulses that allow a controlled interferometric measurement of higher harmonic generation pulses with 12.8 attoseconds in resolution (half the atomic unit of time) and a precision as low as 680 zeptoseconds ($10^{-21}$ seconds). We create two replicas of a driving femtosecond pulse which share the same optical path except at the focus where they converge to two foci. An attosecond pulse train emerges from each focus through the process of HHG. The two attosecond pulse trains from each focus interfere in the far field producing a clear interference pattern in the XUV region. By controlling the relative optical phase between the two driving laser pulses we are able to actively influence the delay between the pulses and are able to perform very stable and precise pump-probe experiments. Because of the phase operation occurs across the entire spatial profile we effectively create two indistinguishable intense laser pulses or a common path interferometer for attosecond pulses. Commonality across the two beams means that they are extremely stable to environmental and mechanical fluctuations up to a Rayleigh range from the focus. In our opinion this represents an ideal source for homodyne and heterodyne spectroscopic measurements with sub-attosecond precision.
\end{abstract}


\section{Introduction}
It takes photons, carrying information about the electromagnetic interaction between electrons and atoms $\approx 0.5$ attoseconds to travel across one carbon atom. This makes attoseconds the natural time scale for bound electrons in atoms, molecules and solids. For this reason attosecond metrology has been at the forefront of the optical sciences for the last fifteen years \cite{Constant1997PRA, paul2001science,Hentschel2001Nature, itatani2002PRL,tzallas2003nature,baltuvska2003nature,kienberger2004nature, Sansone2006Science,corkum2007nature,kolliopoulos2014josab} and it represents the latest time-domain frontier of the quantum world \cite{ossiander2016naturephysics}. Attosecond light pulses are commonly generated using higher-order harmonic generation (HHG) \cite{Harris1993OC,Antoine1996PRL}, a process capable of generating XUV coherent pulses \cite{krause1992PRL,lhullier1993PRL,lewenstein1994PRA}. In addition to generating attosecond pulses, HHG can be used to study the structure and dynamics of atoms and molecules by looking at the spectral content of the XUV spectrum. This approach is known as HHG spectroscopy and it provides a coherent, time-dependent approach to studying structure and dynamics in the quantum world \cite{itatani2004nature,li2008science}. To explore the coherent nature of the method \cite{Mairesse2003science}, several approaches have been used \cite{zhou2008PRL,paul2001science,mairesse2005PRA,kim2013naturephotonics,camper2014PRA,bertrand2013naturephysics} and in this article we will focus mainly on the two-source interferometry method, first proposed in Refs. \cite{Zerne1997PRL} and how our technique is improving the ability to tune the interferometer, while still maintaining attosecond stability.
The basic idea of two-source interferometry with HHG is to generate harmonics from two identical sources or optical foci. The generated XUV pulse trains propagate in space and interfere with each other in the far field, following a pattern that closely resembles a Young's double slit interference pattern. Because the XUV radiation emitted from each source is completely coherent and has a well-defined phase relation with the driving laser, the interference pattern gives details about the phase of the quantum state of the target or more precisely the phase of the induced dipole and electron wave packet. 
\section{Experimental setup}
\renewcommand{\floatpagefraction}{.8}
\begin{figure*}[!t]
	\centering
	\includegraphics[width=0.98\textwidth]{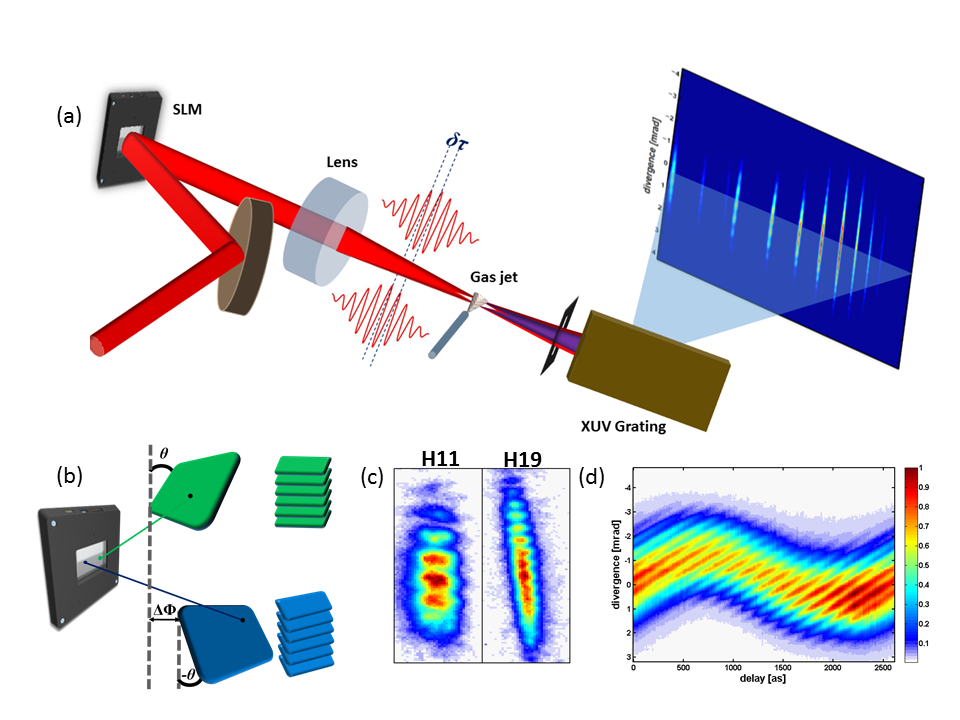}
	\caption{(a) Experimental setup scheme. Light is reflected from an SLM and is focused into a gas jet using a lens (f=50~cm) to produce harmonics from two foci. The harmonics propagate and interfere in the far field (spectrometer detector). (b) Working schematics of the applied phase masks. The two masks are applied to the entire beam in a check pattern. The phases are wrapped in multiples of $2\pi$. (c) Profiles of the 11th and 19th harmonic showing the interference pattern. The number of peaks changes with the wavelength (harmonic order) and with the distance between foci. (d) Evolution of the interference pattern of harmonic 11 as a function of the relative offset phase between the two masks $\Delta \Phi_{SLM}$.}
	\label{fig:Experimental_setup}
\end{figure*}
\begin{figure*}[!t]
	\centering
	\includegraphics[width=0.98\textwidth]{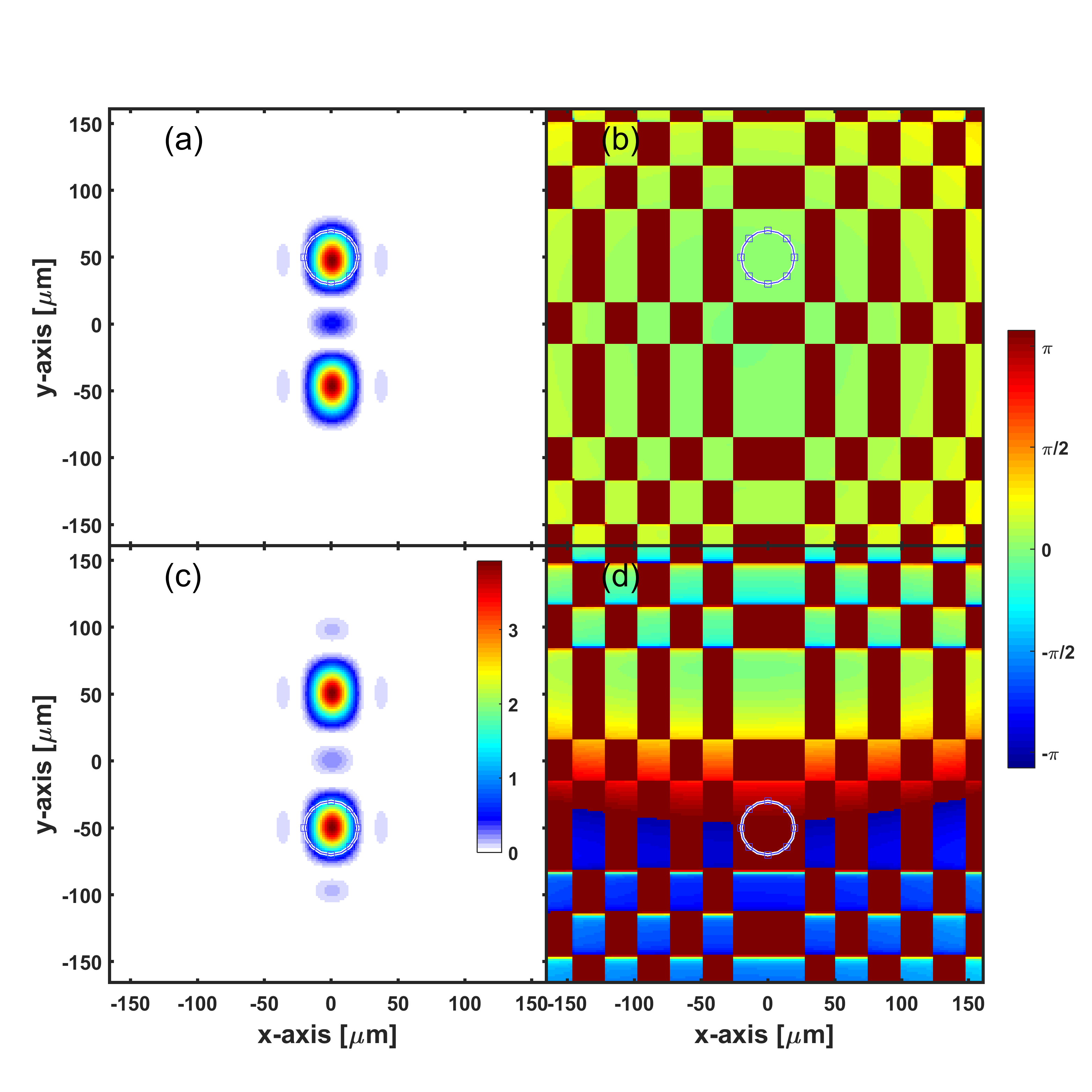}
	\caption{Spatial profiles of the laser at the focus. In panel a), the intensity distribution is given for two laser foci separated by 100$\mu{m}$ and with no additional phase offset, as seen in panel b), where we display the phase distribution of the light.  In panel c), we show the intensity distribution of a phase mask that imprints an phase offset of $\pi/2$ between the two foci. The intensity difference is below 1\% across the beams, when a phase change of $\pi/2$ is applied. The phase distribution in d) does differ between the two foci: A delay of $\pi/2$ is visible between the two foci. Since both foci are fairly close together, we can observe a phase difference of 200~mrad to the desired phase difference, when we are 1 $\sigma$ away from the center of focus (as indicated by the circles)}
	\label{fig:Calc}
\end{figure*}
In this work we generate harmonics from two optical foci ($f_1$ and $f_2$) produced by a system of a two-dimensional spatial light modulator (SLM) and a lens (Fig. \ref{fig:Experimental_setup} a). The two foci are generated using two intertwined phase masks (Fig. \ref{fig:Experimental_setup} b) with opposite wave front tilt. We start with an intense 30~fs pulse incident on the SLM to produce peak intensities at the foci of $\approx 10^{14}$W/cm$^2$. At these intensities, when interacting with a dense gas target, harmonics are generated at each focus. As the harmonics propagate towards the detector they interfere in the far field. An example of the measured interference pattern of harmonics 11th (71.4 nm, 17.4 eV) and 19th (41.3 nm, 30 eV) is shown in Fig. \ref{fig:Experimental_setup} c). Similar interferometric schemes have been used in the past \cite{zhou2008PRL,smirnova2009nature,bertrand2013naturephysics} with HHG, but have to compromise between stability or flexibility. With our method it is possible to control the generation of the one XUV beam (spatially and temporally) independently from the other. This is the novelty of this scheme, which consists an advantage in comparison to previously proposed methods\cite{zhou2008PRL,camper2014PRA,camper2015Photo}. By applying an offset to one mask relative to the other we can control the delay or phase (within one cycle) of one pulse relative to the other. This is simply understood in terms of Fourier optics \cite{voelz2011spie}. Since the SLM is limited in the amount of overall phase ($2\times \pi$ at 785 nm), we wrap the phase similar to Fresnel lenses \cite{voelz2011spie}. Because the mask is a phase change across the entire beam profile, there is only one beam with two diverging wave fronts. With this, we are able to generate two beams that share almost exactly the same beam path. They only become distinguishable at approximately one Rayleigh range before the focus. As the beam is focused in vacuum, the difference in optical path is only dictated by fluctuations in the background pressure over a distance of a few millimeters. The method of pulse shaping has been actively used as a means of generating zeptosecond precision interferometers \cite{kohler2011OE} in a frequency domain setup, which does not allow the generation of two spatially separated pulses.
It is important to remark that the phases are applied to the entire beam and not split in distinguishable halves, like in previous works \cite{camper2015Photo,zhou2008PRL}. The phase and tilt of each pulse is identified by using a checkerboard phase pattern where $f_1$ is controlled by ``white" squares and $f_2$ is controlled by ``black" squares. The main advantage of using an intertwined pattern is that the two beams are completely indistinguishable from each other up to a few millimeters from the focus, which was also the case in previous methods \cite{camper2015Photo,zhou2008PRL}. Nevertheless, with the proposed scheme, the sampling on the laser mode is much better, the alignment process much easier and the whole optics setup more resistant to slight beam pointing instabilities. In addition, the distance in between the two foci (the two secondary sources) can be tuned dynamically, without any modification on the optical setup and without any realignment, only by applying on the SLM a new set of sub-patterns with a different tilt with each other. This possibility can go so far as to isolate one of the two sources. One has just to increase the tilt of the sub-pattern which corresponds to the second source so much as to transpose this last one far away from the target (for an application which makes use of this, see section 4). Consequently, it is possible to check the emission from each one of the two sources separately before letting them interfere in the far field. To our knowledge this is a completely new possibility in the field of HHG spectroscopy which can be utilized in order to in situ check if the two secondary sources are indeed identical. To preserve phase matching in HHG the Rayleigh range is chosen to be much larger than the interacting region ($500~\mu$m in our case). Also, the distance between the two foci is smaller than the inner diameter of the glass capillary used to generate the gas jet. Practically, this means that differences in the optical path will be dictated by changes in pressures of $10^{-4}$ to $10^{-5}$ Torr over one Rayleigh range. 

\begin{figure}
	\centering
	\includegraphics[width=0.9\columnwidth]{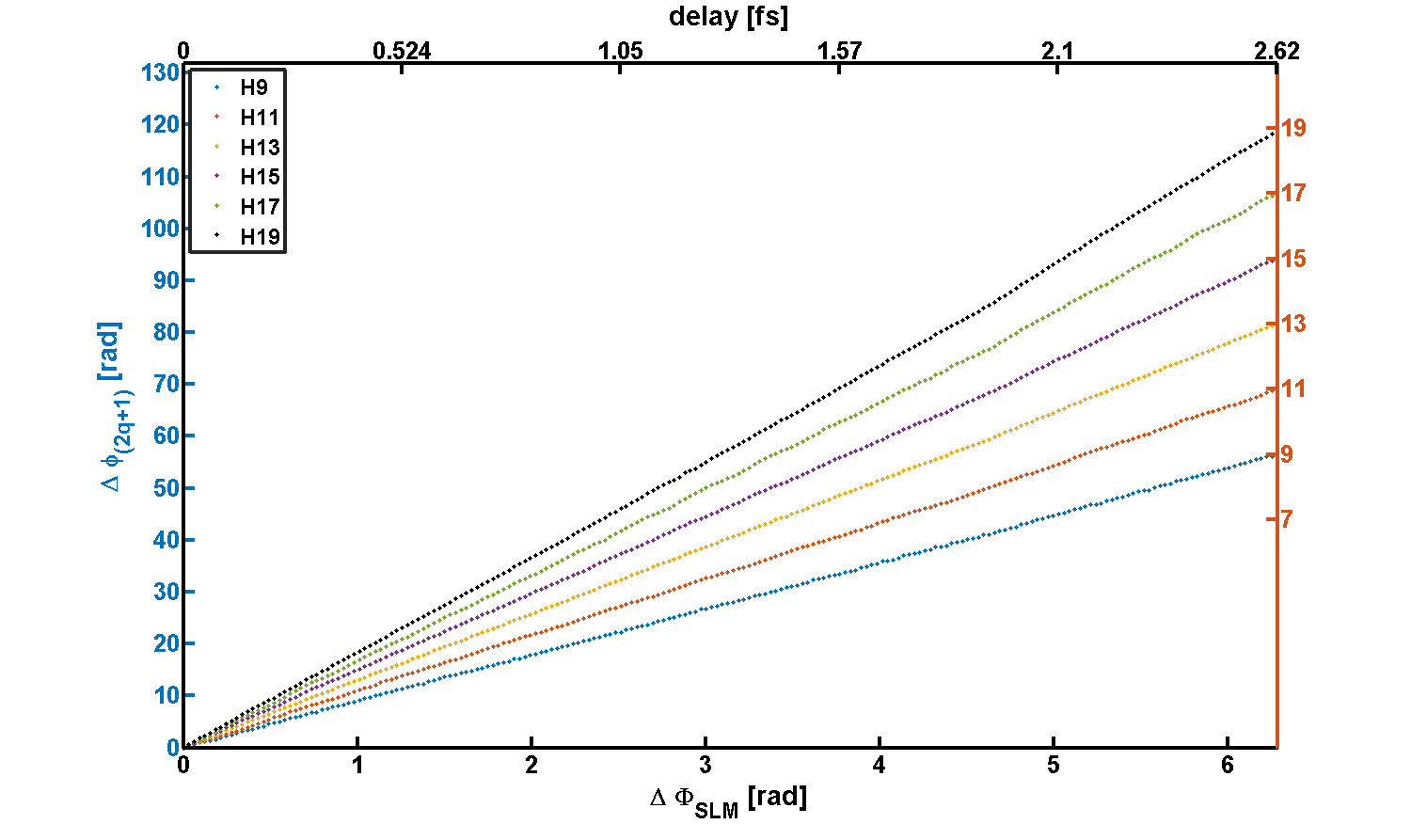}
	\caption{Phase of harmonics 9 to 19 (left axis) measured from the interferograms as shown in Fig. \ref{fig:Experimental_setup} c) as a function of the relative SLM phase $\Delta \Phi_{SLM}$ (bottom axis). The right axis is the phase in units of $2\pi$ showing that the phase evolution of each harmonic $2q+1$ is $\Delta \phi_{(2q+1)} = (2q+1) \times \Delta \Phi_{SLM}$. Top axis is the corresponding delay time for a 785 nm wavelength pulse. The SLM has access to 205 phase values thus capable of a resolution of 12.8 attoseconds.}
	\label{fig:HHG_phase}
\end{figure}
\section{Stability and precision measurement}
The overall phase of the driving field has a great impact on the generation of harmonics \cite{balcou1997PRA}. In Figure \ref{fig:Calc}, we show that our tilted Fresnel masks produce Gaussian shaped foci with transversal phase distributions which are almost flat. In the figure and with a separation of 100~$\mu{m}$, the two laser fields still overlap slightly in the focus and the spatial phase of the two sources is altered (through interference effects). The intensity distribution is shown in the left panel. There, intensity differences of less than 1\% can be observed between phase masks of altered delay. In the middle and right panel, we show the spatial phase distribution of two separate masks. There, the phase profile across the foci is flat, except at the wings (we define wings as intensities of less than 10\% of the peak intensity) for a few values of the relative phase between the two foci(e.g. $pi/4$ and $3\pi/4$). In this case we can observe a slight wave front tilt of the driving field which is responsible for the observed characteristic "S-shape" in Figure \ref{fig:Experimental_setup} panel (d). An improvement could take place by increasing the distance in between the two foci (simply by increasing the tilt of the two sub-patterns applied on the SLM). Unfortunately, at the same time, the fringe spacing at the detector plane is getting reduced. We control the relative phase between the two foci by adding an offset phase to the one of the tilted Fresnel masks (see Fig. \ref{fig:Experimental_setup} b)). Since this induces a relative phase difference between the two masks at the SLM, we call this $\Delta \Phi_{SLM}$. The evolution of the interference fringes in harmonic 11 as a function of $\Delta \Phi_{SLM}$ is shown in Fig. \ref{fig:Experimental_setup} d). In our SLM we can control $\Delta \Phi_{SLM}$ with a step size of $\approx 12.8$ attoseconds at the center wavelength of 785 nm. We should point out that because the two foci are exact replicas of each other and are generated from the same driving pulse, the relative delay between of the two sources is completely independent of any fluctuation of the laser. This relative delay, through the HHG process \cite{lewenstein1994PRA}, is inherited to the relative phase of the harmonic fields emanating from each one of the two foci, multiplied by the order of the harmonic. We have to notice here that since the two foci are identical, the term of the atomic phase (which depends on the driving field intensity)\cite{lewenstein1994PRA}, is the same for both of them. Consequently, it does not affect their phase difference.  

\begin{figure}
\centering
\includegraphics[width=0.65\columnwidth]{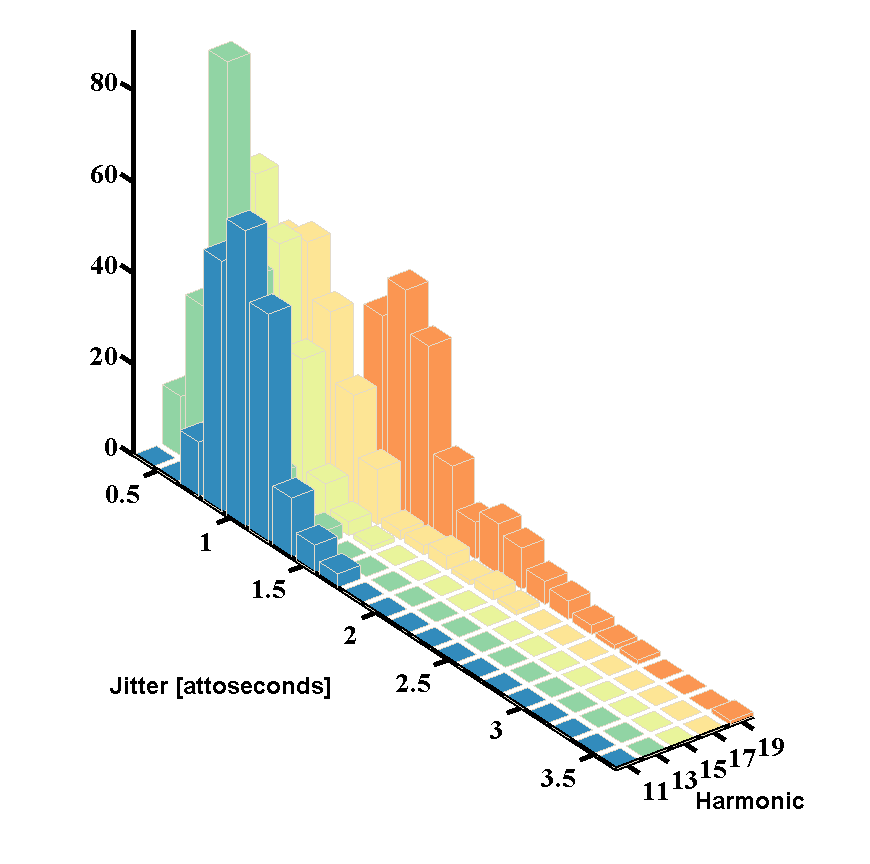}
\caption{Histogram of the jitter for all delays as a function of harmonic. The jitter is calculated as the standard error, for each delay value taking into account the 17 measurements done at each delay. Each of the 17 measurements are made up 600 laser shots.}
\label{fig:histogram}
\end{figure}

To more quantitatively extract the relative phase between the two harmonic sources $\Delta \phi_q$, as a function of the relative delay of the two foci we need to calculate the phase evolution for each harmonic. We can do this calculation in two ways. One option is to measure the relative displacement of the maxima and minima in the fringe pattern keeping in mind that a $2\pi$ shift corresponds to the distance between two peaks \cite{Zerne1997PRL}. Our method makes use of Fourier analysis of the fringe pattern. In this approach $\Delta \phi_q$ is the phase of the fringe frequency of the Fourier spectrum, which corresponds to the inverse spacing between peaks in the detector. Because the fringe separation in the detector is proportional to $\lambda_{(2q+1)}/D$ with D the spacing between the two sources, the fringe spacing reduces with harmonic order. For this study, we kept the separation between the two foci at $100 \mu$m. The relative phase evolution for harmonics 9 to 19 (left axis) using the Frequency Fourier analysis is shown in Fig. \ref{fig:HHG_phase}. In addition, we show on the right axis the phase in units of $2\pi$ for each harmonic. From this plot the scaling of the phase $\Delta \phi_{(2q+1)}$, for each harmonic of order $2q+1$, as a function of $\Delta \Phi_{SLM}$ is clear.  This scaling is another demonstration of the self-balancing geometry of our interferometer. For each oscillation period of the fundamental there are exactly ${(2q+1)}$ oscillations in the electromagnetic field for harmonic ${(2q+1)}$. Again, we are able to observe this thanks to the self-referencing nature of the interferometer. To be able to observe oscillations in the harmonics we of course need attosecond resolution in our time steps. The top axis in Fig. \ref{fig:HHG_phase} shows the corresponding delay time for a 785 nm center wavelength pulse. As mentioned before the SLM has access to 205 effective phase values thus capable of a resolution of 12.8 attoseconds. This resolution was confirmed by measuring the amount the fringes moved for each harmonic. However, the main limitation in most interferometers is not in the resolution but rather in the precision. Temperature, humidity and mechanical instabilities are often enough to cause errors in the orders of hundreds of attoseconds. To further demonstrate the power of our method, we calculate the jitter for harmonics 9 to 19 using 17 independent experiments at each delay. The jitter is calculated from the standard error of the phase using the 17 measurements. Each one of the 17 measurements is comprised of 600 lasers shots, the number of laser shots chosen to form a harmonic picture. Figure \ref{fig:histogram} shows the histogram of the jitter values for the 205 delay values for harmonics 11 to 19. The results are plotted as a histogram of the jitter values. Harmonic 9 is not included in the figure but is included in the summary below.

\begin{table}[]
	\begin{center}
		\begin{tabular}{  l | l | l }

			HO & SE [as] & deviation [as]  \\ \hline
			9 & 3.10 & 2.8  \\ \hline
			11 & 0.80 & 0.72  \\ \hline
			13 & 0.68 & 0.60  \\ \hline
			15 & 0.71 & 0.63  \\ \hline
			17 & 0.80 & 0.71  \\ \hline
			19 & 1.10 & 0.94  \\ \hline
		\end{tabular}
		\caption{Column ``SE" are the standard error estimates calculated from 17 images taken for each harmonic with each image consisting of 600 laser shots. Since we can assume uncorrelated errors for the temporal jitter, we can also use the 205 delay ``experiments" as  measurements for the error.}
		\label{tab:errors}
	\end{center}
\end{table}

The results from Fig. \ref{fig:histogram} are summarized in Table \ref{tab:errors}. The SE column reports the standard error of the jitter from the histogram and the deviation column is the standard deviation from the same distribution. From both, the histogram and the table, we can clearly observe that we have a precision better than one attosecond. Harmonic 13 has the best precision of 680 zeptoseconds followed by harmonic 15 with a precision of 710 zeptoseconds. We should mention that harmonic 9 and 19 have the worst resolution of 3.1 and 1.1 attoseconds. These larger uncertainties are due to poor signal levels in harmonic 9 and due to a lack of resolution in the fringes in harmonic 19. The low signal of harmonic 9 is due to the response of the holographic grating used. As mentioned, the resolution of the fringes can be improved by changing the distance between the foci.\newline

\begin{figure}
    \centering
    \includegraphics[width=0.9\columnwidth]{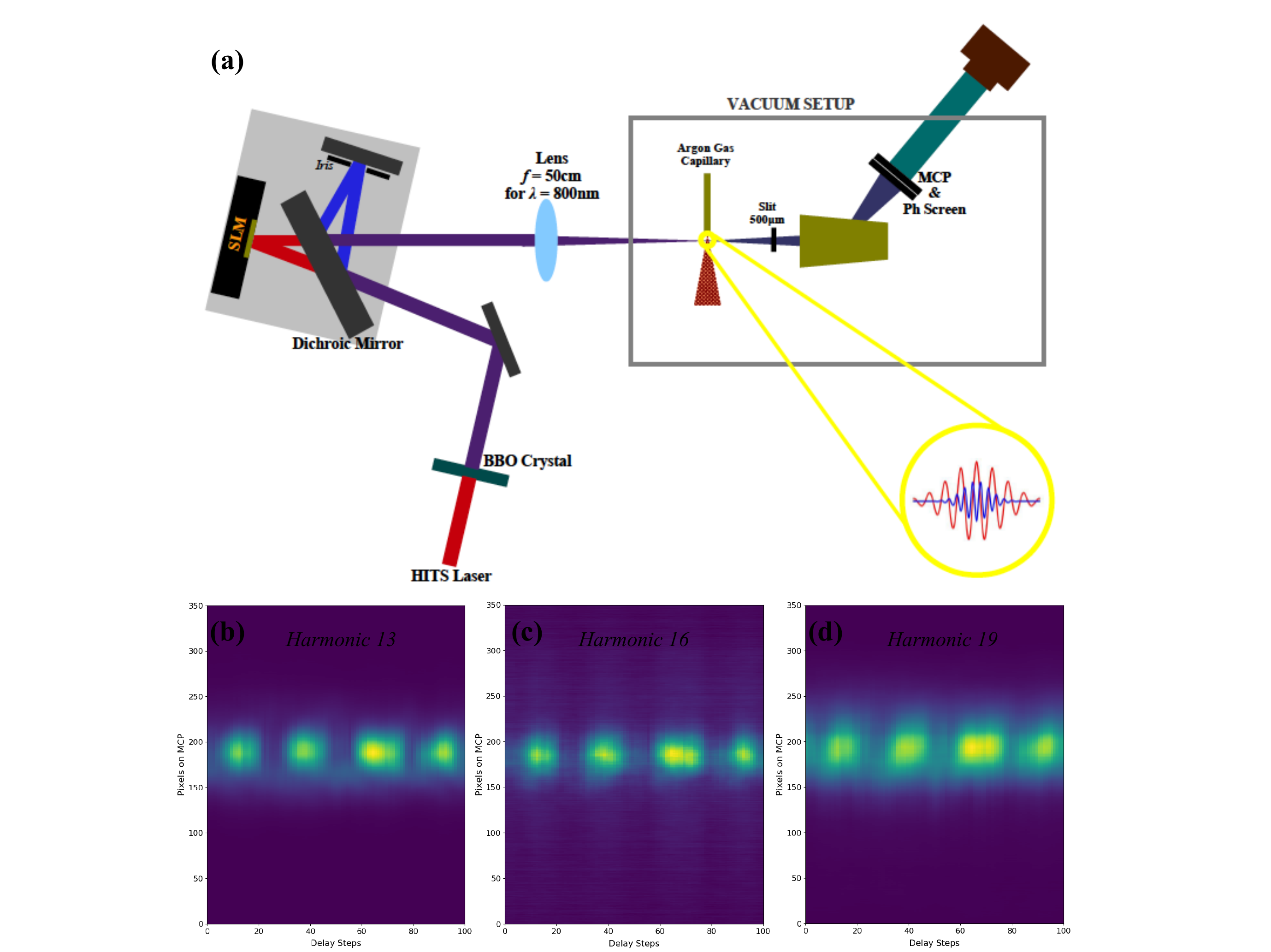}
    \caption{(a) Experimental setup scheme. The 2nd harmonic of the 785 nm light is generated in a BBO crystal. 785 nm and 392 nm beams are separated by a dichroic mirror. The IR part is reflected from an SLM and its 2nd harmonic is reflected from a silver mirror. In the SLM masks is now added a Fresnel lens pattern (f = 11 m). After the recombination, 785 nm and 393 nm beams are focused together into a gas jet at a single focus. (b), (c) and (d) Evolution of the harmonics 13, 16 and 19 as a function of the offset phase applied on the SLM sub-pattern. This corresponds to a tunable phase-delay between the 785 nm and the 393 nm electric fields.}
\label{fig:fig4-IR & Blue}
\end{figure}

\section{A two-color Michelson-type interferometer}
We have already discussed on the possibility to use the above scheme as a self-referencing interferometer. Here we will show that the phase delay induced by our method can be considered as well in respect to an external field. In other words, we will show that, with our scheme, the SLM can successfully substitute the movable arm of a Michelson-type interferometer, that the SLM is able to perfectly act as a delay stage of high precision and stability. 

Here we retrieve the delay dependence on HHG, when driven by a two color laser field in argon gas.
The experimental setup is shown in in Figure \ref{fig:fig4-IR & Blue} panel(a), where an incoming 785 nm laser field produces
its second harmonic in a 200 $\mu$m thick beta barium borate (BBO) crystal. A dichroic mirror separates the 393 nm radiation from the 785 nm one. The former is reflected on a flat silver mirror and the last is reflected on the SLM surface. They recombine and they are both focused on the target. On the SLM is applied a series of masks which induce in the gas target area a single focal spot identical to the one of the pair of spots used before. This is feasible simply by increasing by one order of magnitude the tilt angle of one of the subpatterns. As a result, the second focal spot is formed far away from the gas-target. The two components (785 nm and 393 nm) are focused using a lens with a focal length equal to 50 cm for a wavelength equal to 800 nm. The focal length of the lens is not the same for the 393 nm centered light. We are able to bring the two beams to focus together by simply adding a fresnel lens pattern (with focal length equal to 11 m) on the top of each one of the masks that we apply on the SLM. With this small modification, in the series of masks, and not in the experimental setup, the IR is brought to focus together with its 2nd harmonic on the gas-target. This proves the high versatility of the proposed method. The polarizations of the IR and its 2nd harmonic are perpendicular to each other and the amplitude ratio of the 393 nm field in respect to the 785 nm field is 0.22. As demonstrated in numerous previous works \cite{Brugnera2010OL,Kim2005PRL}, the second harmonic can act as a gate on the process of high harmonic generation. Depending on the delay between 785 and 392 nm, high-Harmonic generation of any specific order is either suppressed or enhanced. This appears in the characteristic form of yield oscillations (see Figure \ref{fig:fig4-IR & Blue} panels(b)-(d)), as it has been confirmed before.

\section{Summary}
To conclude, we have presented in this work a novel interferometer, for trains of attosecond pulses generated through HHG, with tunable arms and high stability. An impressive delay resolution with high precision has been proven by the fact that the phase difference of each harmonic order ${(2q+1)}$, $\phi_{(2q+1)}$, scales with the phase of the fundamental as $\phi_{(2q+1)}= {(2q+1)}\times \phi_{fundamental}$. 
This interferometer resembles a common path homodyne detector for the fundamental as two identical copies of the driving pulse are created at the focus by using an SLM and a lens. Since the two beams are indistinguishable until $\approx$ one Rayleigh range from the focus, their optical path are identical and almost jitter free. Because of this, the phase between the two foci is perfectly locked. For the first time, the relative phases of the driving field in each one of the two foci can be manipulated. The trains of attosecond pulses, which emanate from these secondary sources, can be delayed with respect to each other with a resolution of 12.8 attoseconds (half of the atomic unit of time). This delay can be controlled with a precision of 680 zeptoseconds for harmonic 13.

\section*{Funding}
This work was supported by the Chemical Sciences, Geosciences, and Biosciences Division, Office of Basic Energy Sciences, Office of Science, U.S. Department of Energy (DOE) under Grant No. DE-FG02-86ER13491.

\bibliography{attosecond_interferometer_V15.bib}

\begin{thebibliography}{10}
\newcommand{\enquote}[1]{``#1''}

\bibitem{Constant1997PRA}
E.~Constant, V.~D. Taranukhin, A.~Stolow, and P.~B. Corkum, \enquote{{Methods
  for the measurement of the duration of high-harmonic pulses},} Physical
  Review A \textbf{56}, 3870--3878 (1997).

\bibitem{paul2001science}
P.~.~M. Paul, E.~Toma, P.~Breger, G.~Mullot, F.~Aug{\'e}, P.~Balcou, H.~Muller,
  and P.~Agostini, \enquote{Observation of a train of attosecond pulses from
  high harmonic generation,} Science \textbf{292}, 1689--1692 (2001).

\bibitem{Hentschel2001Nature}
M.~Hentschel, R.~Kienberger, C.~Spielmann, G.~a. Reider, N.~Milosevic,
  T.~Brabec, P.~Corkum, U.~Heinzmann, M.~Drescher, and F.~Krausz,
  \enquote{{Attosecond metrology.}} Nature \textbf{414}, 509--13 (2001).

\bibitem{itatani2002PRL}
J.~Itatani, F.~Qu{\'e}r{\'e}, G.~L. Yudin, M.~Y. Ivanov, F.~Krausz, and P.~B.
  Corkum, \enquote{Attosecond streak camera,} Physical Review Letters
  \textbf{88}, 173903 (2002).

\bibitem{tzallas2003nature}
P.~Tzallas, D.~Charalambidis, N.~Papadogiannis, K.~Witte, and G.~D. Tsakiris,
  \enquote{Direct observation of attosecond light bunching,} Nature
  \textbf{426}, 267--271 (2003).

\bibitem{baltuvska2003nature}
A.~Baltu{\v{s}}ka, T.~Udem, M.~Uiberacker, M.~Hentschel, E.~Goulielmakis,
  C.~Gohle, R.~Holzwarth, V.~Yakovlev, A.~Scrinzi, T.~W. H{\"a}nsch
  \emph{et~al.}, \enquote{Attosecond control of electronic processes by intense
  light fields,} Nature \textbf{421}, 611--615 (2003).

\bibitem{kienberger2004nature}
R.~Kienberger, E.~Goulielmakis, M.~Uiberacker, A.~Baltuska, V.~Yakovlev,
  F.~Bammer, A.~Scrinzi, T.~Westerwalbesloh, U.~Kleineberg, U.~Heinzmann
  \emph{et~al.}, \enquote{Atomic transient recorder,} Nature \textbf{427},
  817--821 (2004).

\bibitem{Sansone2006Science}
G.~Sansone, E.~Benedetti, F.~Calegari, C.~Vozzi, L.~Avaldi, R.~Flammini,
  L.~Poletto, P.~Villoresi, C.~Altucci, R.~Velotta, S.~Stagira, S.~{De
  Silvestri}, and M.~Nisoli, \enquote{{Isolated Single-Cycle Attosecond
  Pulses},} Science \textbf{314} (2006).

\bibitem{corkum2007nature}
P.~{\'a}. Corkum and F.~Krausz, \enquote{Attosecond science,} Nature Physics
  \textbf{3}, 381--387 (2007).

\bibitem{kolliopoulos2014josab}
G.~Kolliopoulos, P.~Tzallas, B.~Bergues, P.~Carpeggiani, P.~Heissler,
  H.~Schr{\"o}der, L.~Veisz, D.~Charalambidis, and G.~D. Tsakiris,
  \enquote{Single-shot autocorrelator for extreme-ultraviolet radiation,}
  Journal of the Optical Society of America B \textbf{31}, 926--938 (2014).

\bibitem{ossiander2016naturephysics}
M.~Ossiander, F.~Siegrist, V.~Shirvanyan, R.~Pazourek, A.~Sommer, T.~Latka,
  A.~Guggenmos, S.~Nagele, J.~Feist, J.~Burgd{\"o}rfer \emph{et~al.},
  \enquote{Attosecond correlation dynamics,} Nature Physics  (2016).

\bibitem{Harris1993OC}
S.~E. {Harris}, J.~J. {Macklin}, and T.~W. {H{\"a}nsch}, \enquote{{Atomic scale
  temporal structure inherent to high-order harmonic generation},} Optics
  Communications \textbf{100}, 487--490 (1993).

\bibitem{Antoine1996PRL}
P.~Antoine, A.~L'Huillier, and M.~Lewenstein, \enquote{Attosecond pulse trains
  using high-order harmonics,} Physical Review Letters \textbf{77}, 1234--1237
  (1996).

\bibitem{krause1992PRL}
J.~L. Krause, K.~J. Schafer, and K.~C. Kulander, \enquote{High-order harmonic
  generation from atoms and ions in the high intensity regime,} Physical Review
  Letters \textbf{68}, 3535 (1992).

\bibitem{lhullier1993PRL}
A.~L’Huillier and P.~Balcou, \enquote{High-order harmonic generation in rare
  gases with a 1-ps 1053-nm laser,} Physical Review Letters \textbf{70}, 774
  (1993).

\bibitem{lewenstein1994PRA}
M.~Lewenstein, P.~Balcou, M.~Y. Ivanov, A.~L’huillier, and P.~B. Corkum,
  \enquote{Theory of high-harmonic generation by low-frequency laser fields,}
  Physical Review A \textbf{49}, 2117 (1994).

\bibitem{itatani2004nature}
J.~Itatani, J.~Levesque, D.~Zeidler, H.~Niikura, H.~P{\'e}pin, J.-C. Kieffer,
  P.~B. Corkum, and D.~M. Villeneuve, \enquote{Tomographic imaging of molecular
  orbitals,} Nature \textbf{432}, 867--871 (2004).

\bibitem{li2008science}
W.~Li, X.~Zhou, R.~Lock, S.~Patchkovskii, A.~Stolow, H.~C. Kapteyn, and M.~M.
  Murnane, \enquote{Time-resolved dynamics in n2o4 probed using high harmonic
  generation,} Science \textbf{322}, 1207--1211 (2008).

\bibitem{Mairesse2003science}
Y.~Mairesse, A.~de~Bohan, L.~J. Frasinski, H.~Merdji, L.~C. Dinu,
  P.~Monchicourt, P.~Breger, M.~Kova{\v c}ev, R.~Ta{\"\i}eb, B.~Carr{\'e},
  H.~G. Muller, P.~Agostini, and P.~Sali{\`e}res, \enquote{Attosecond
  synchronization of high-harmonic soft x-rays,} Science \textbf{302},
  1540--1543 (2003).

\bibitem{zhou2008PRL}
X.~Zhou, R.~Lock, W.~Li, N.~Wagner, M.~M. Murnane, and H.~C. Kapteyn,
  \enquote{Molecular recollision interferometry in high harmonic generation,}
  Physical Review Letters \textbf{100}, 073902 (2008).

\bibitem{mairesse2005PRA}
Y.~Mairesse and F.~Qu{\'e}r{\'e}, \enquote{Frequency-resolved optical gating
  for complete reconstruction of attosecond bursts,} Physical Review A
  \textbf{71}, 011401 (2005).

\bibitem{kim2013naturephotonics}
K.~T. Kim, C.~Zhang, A.~D. Shiner, B.~E. Schmidt, F.~L{\'e}gar{\'e},
  D.~Villeneuve, and P.~Corkum, \enquote{Petahertz optical oscilloscope,}
  Nature Photonics \textbf{7}, 958--962 (2013).

\bibitem{camper2014PRA}
A.~Camper, T.~Ruchon, D.~Gauthier, O.~Gobert, P.~Sali{\`e}res, B.~Carr{\'e},
  and T.~Auguste, \enquote{High-harmonic phase spectroscopy using a binary
  diffractive optical element,} Physical Review A \textbf{89}, 043843 (2014).

\bibitem{bertrand2013naturephysics}
J.~Bertrand, H.~W{\"o}rner, P.~Sali{\`e}res, D.~Villeneuve, and P.~Corkum,
  \enquote{Linked attosecond phase interferometry for molecular frame
  measurements,} Nature Physics \textbf{9}, 174--178 (2013).

\bibitem{Zerne1997PRL}
R.~Zerne, C.~Altucci, M.~Bellini, M.~B. Gaarde, T.~W. H{\"{a}}nsch,
  A.~L'Huillier, C.~Lyng{\aa}, and C.-G. Wahlstr{\"{o}}m,
  \enquote{{Phase-Locked High-Order Harmonic Sources},} Physical Review Letters
  \textbf{79}, 1006--1009 (1997).

\bibitem{smirnova2009nature}
O.~Smirnova, Y.~Mairesse, S.~Patchkovskii, N.~Dudovich, D.~Villeneuve,
  P.~Corkum, and M.~Y. Ivanov, \enquote{High harmonic interferometry of
  multi-electron dynamics in molecules,} Nature \textbf{460}, 972--977 (2009).

\bibitem{camper2015Photo}
A.~Camper, A.~Ferr{\'e}, N.~Lin, E.~Skantzakis, D.~Staedter, E.~English,
  B.~Manschwetus, F.~Burgy, S.~Petit, D.~Descamps \emph{et~al.},
  \enquote{Transverse electromagnetic mode conversion for high-harmonic
  self-probing spectroscopy,} in \enquote{Photonics,} , vol.~2
  (Multidisciplinary Digital Publishing Institute, 2015), vol.~2, pp. 184--199.

\bibitem{voelz2011spie}
D.~G. Voelz, \emph{Computational fourier optics: a MATLAB tutorial} (Spie Press
  Bellingham, Wash, USA, 2011).

\bibitem{kohler2011OE}
J.~K{\"o}hler, M.~Wollenhaupt, T.~Bayer, C.~Sarpe, and T.~Baumert,
  \enquote{Zeptosecond precision pulse shaping,} Optics Express \textbf{19},
  11638--11653 (2011).

\bibitem{balcou1997PRA}
P.~Balcou, P.~Salieres, A.~L'Huillier, and M.~Lewenstein, \enquote{Generalized
  phase-matching conditions for high harmonics: The role of field-gradient
  forces,} Physical Review A \textbf{55}, 3204 (1997).

\bibitem{Brugnera2010OL}
L.~Brugnera, F.~Frank, D.~J. Hoffmann, R.~Torres, T.~Siegel, J.~G. Underwood,
  E.~Springate, C.~Froud, E.~I.~C. Turcu, J.~W.~G. Tisch, and J.~P. Marangos,
  \enquote{Enhancement of high harmonics generated by field steering of
  electrons in a two-color orthogonally polarized laser field,} Optics Letters
  \textbf{35}, 3994--3996 (2010).

\bibitem{Kim2005PRL}
I.~J. Kim, C.~M. Kim, H.~T. Kim, G.~H. Lee, Y.~S. Lee, J.~Y. Park, D.~J. Cho,
  and C.~H. Nam, \enquote{Highly efficient high-harmonic generation in an
  orthogonally polarized two-color laser field,} Physical Review Letters
  \textbf{94} (2005).

\end{thebibliography}
\end{document}